\documentclass[useAMS]{mn2e}
\usepackage{times}
\usepackage{epsfig}
\def\gtrsim{\mathrel{\hbox{\rlap{\hbox{\lower4pt\hbox{$\sim$}}}\hbox{$>$}}}}

\def \etal   {\hbox{\it et~al.\/}}

\title[Polarization during caustic crossings]
{The polarisation signature from microlensing of circumstellar envelopes in
caustic crossing events }
\author[R. Ignace, J. E. Bjorkman, and H. M. Bryce]
{
R.~Ignace$^{1,2}$, 
J.~E.~Bjorkman$^{3}$, and
H.~M.~Bryce$^{1,4}$ \\
$^1$ Department of Astronomy, University of Wisconsin, 475 North
Charter Street, Madison, WI  53706, USA\\
$^2$ Department of Physics, Astronomy, \& Geology, East Tennessee State
University, Johnson City, TN  37614, USA\\
$^3$ Ritter Observatory, Department of Physics and Astronomy,
University of Toledo, Toledo, OH 43606, USA\\
$^4$ Department of Physics and Astronomy, University of Glasgow,
Glasgow, Scotland, UK\\
}

\begin{document}
\maketitle

\begin{abstract}

In recent years it has been shown that microlensing is a powerful
tool for examining the atmospheres of stars in the Galactic Bulge and
Magellanic Clouds. The high gradient of magnification across the source
during both small impact parameter events and caustic crossings offers
a unique opportunity for determining the surface brightness profile
of the source. Furthermore, models indicate that these events can also
provide an appreciable polarisation signal -- arising from differential
magnification across the otherwise symmetric source. Earlier work
has addressed the signal from a scattering photosphere for both point
mass lenses and caustic crossings.  In a previous paper, polarimetric
variations from point lensing of a circumstellar envelope were considered,
as would be suitable for an extended envelope around a red giant. In this
work we examine the polarisation in the context of caustic crossing
events, the scenario which represents the most easily accessible
situation for actually observing a polarisation signal in Galactic
microlensing. Furthermore we present an analysis of the effectiveness
of using the polarimetric data to determine the envelope properties,
illustrating the potential of employing polarimetry in addition to
photometry and spectroscopy with microlensing follow-up campaigns.

\end{abstract}


\section{Introduction}

Much attention has been paid to the situation in which the analytic
case of the magnification of a point source by a point lens
breaks down. It has been noted that for small lens--source separations,
the finite size of the source star needs to be considered
(Nemiroff \& Wickramasinghe 1994; Witt \& Mao 1994; Witt 1995;
Peng 1997). Such events not only constrain the lens properties by
breaking the degeneracy in the event parameters (Gould 1994) but also
provide valuable stellar atmosphere information, such as limb
darkening (Valls-Gabaud 1998; Hendry et al.\ 1998), polarisation
(Simmons et al.\ 1995a,b; Newsam et al.\ 1998), motions in circumstellar
envelopes (Ignace \& Hendry 1999) and the presence of starspots
(Heyrovsk\'{y} \& Sasselov 2000). The opportunity for studying
stellar atmospheres through microlensing as described in Sackett
(2001) and Gould (2001) has clear advantages over other methods
such as eclipsing binaries, because the source and means of studying
the source are not coupled and the flux from the source is magnified
rather than diminished.

Binary lenses produce another deviation from a standard microlensing
lightcurve, because caustics are produced (e.g., Schneider \& Weiss
1986). It is the high gradient of magnification across the caustic
that allows one to infer information about the source intensity
profile, even though the source images are not individually resolvable
with current instruments.  Despite binary lens events only accounting for
about 5\% of microlensing events (e.g., Alcock \etal\ 2000), it is more
likely that finite source effects will be relevant for a binary lens
rather than a point lens.  The reason is that finite source effects
are mainly discernible only when the lens transits the source itself
(Gould 1994).  Since the angular Einstein radius $\theta_{\rm E}$ (see
Eq.~\ref{eq:2}) is usually much larger than the angular source size,
such transits tend to be rare.  On the other hand, the caustic structures
giving rise to high magnifications in binary lens events are spatially
extended (of order $\theta_{\rm E}$ in scale), and so caustic crossings
by the more distant source are relatively common in events associated
with binary lenses.  So it is more expedient to study the resolution
of stellar atmospheres due to binary lenses than point lenses. The
structure of the caustics produced and the resulting magnification and
lightcurves from binary microlenses has been discussed extensively in
the literature (Mao \& Paczy\'{n}ski 1991; Mao \& Di Stefano 1995; Di
Stefano \& Perna 1997; Dominik 1998; Gaudi \& Gould 1999; Dominik 2004b).
A number of papers have reported on source properties (such as limb
darkening) derived from the light curves of binary lens events (Albrow
\etal\ 1999; Afonso \etal\ 2000; Albrow \etal\ 2000; Albrow \etal\ 2001a;
Albrow \etal\ 2001b; Abe \etal\ 2003; Cassan \etal\ 2004).

Considerable theoretical work on the polarisation signatures from
microlensing has also been carried out. Schneider \& Weiss (1986)
were the first to discuss the use of caustic crossings for inferring
source intensity profiles.  Schneider \& Wagoner (1987) calculated
the polarisation from the lensing of thick scattering photospheres
of supernovae during such crossings. Simmons et al.\
(1995a,b) examined the polarisation from an electron scattering
atmosphere being microlensed by a point mass, thus allowing
the limb polarisation to be measured. Newsam et al.\ (1998)
used this analysis to demonstrate that even relatively `poor'
polarimetric data can considerably improve the determination of
stellar radii.  In turn, Agol (1996) modelled the polarisation from
an electron scattering stellar atmosphere by a binary lens. More recently,
Simmons et al.\ (2002) examined the polarisation from an
extended envelope for a point mass lens. The work of this paper 
applies the same atmosphere model as in Simmons et al.\ (2002)
to the case of caustic crossings -- arguably the `best case
scenario' for probing the extended envelope.  

The structure of this paper is as follows. In Section 2 we discuss
the microlensing of extended sources.  In Section 3 we describe the
polarisation intensity maps used in the calculation of the microlensing
lightcurves. In Section 4 we provide representative lightcurves for
both the polarisation and flux signals for a range of source
parameters.  We also discuss the observational implications of our
results.  The duration of caustic crossing events typically occurs
over one night, so we believe that it is important to explore
different observing strategies to find the most suitable way of
observing these events while effectively recovering the source
and envelope properties relative to the Einstein radius.
Concluding remarks are presented in Section 5.

\section{Microlensing of extended sources}

The magnification of a point source by a point lens
is given by 

\begin{equation}
A_{\rm pt} =\frac{u^{2}+2}{u\sqrt{u^{2}+4}}
\label{eq:1}
\end{equation}

\noindent where $u$ is the angular separation between the lens and source
as normalized by the angular Einstein radius $\theta_{\rm E}$, which is
given by

\begin{equation}
\theta_{\rm E} = \sqrt{\frac{4GM_{\rm L}}{c^{2}}\frac{(D_{\rm S}-D_{\rm L})}{D_{\rm L}D_{\rm S}}},
\label{eq:3}
\end{equation}

\noindent where $M_{\rm L}$, $D_{\rm S}$, and $D_{\rm L}$ are the lens
mass, the distance to source and the distance to lens respectively.
The lensing is a transient event with $u=u(t)$ a function of time
given by

\begin{equation}
u(t)=\sqrt{u_0^{2}+\frac{(t-t_{0})^{2}}{t_{\rm E}^{2}}},
\label{eq:2}
\end{equation}

\noindent where $t_{0}$ is the time of maximum magnification at $u_0$
the impact parameter.
The parameter $t_{\rm E}$ is the crossing time of $\theta_{\rm E}$:

\begin{equation}
t_{\rm E}=\theta_{\rm E}/\mu_{\rm rel},
\label{eq:4}
\end{equation}

\noindent where $\mu_{\rm rel}$ is the relative proper motion between
the lens and the source.

The point source approximation is valid when $\theta_{\rm S} \ll
\theta_{\rm E}$, for $\theta_{\rm S}$ the angular radius of the source
(which we shall assume is circularly symmetric).  When this no longer
holds -- such as when the lens transits the source -- the magnification is
determined by an intensity-weighted integral over the projected surface
of the source. The resultant is,

\begin{equation}
A_{\rm net}(t) =
\frac{\int_{0}^{2\pi}\int_{0}^{\theta_{\mathrm{S}}}I(\theta,\alpha)A_{\rm pt}
	(u(\theta_L,\theta,\alpha,t))\,\theta\, d\theta\, d\alpha} 
{\int_{0}^{2\pi}\int_{0}^{\theta_{\mathrm{S}}}I(\theta,\alpha)\theta\,d\theta
	\,d\alpha}
\label{eq:5}
\end{equation} 

\noindent where $\theta$ is an angular radius measured from the source
origin, and $\alpha$ is an azimuthal angle about that origin.  The angle
$\theta_{\rm L}$ is the angular separation between the point lens and
the source centre.  The integral means that different emitting elements
of the projected source make weighted contributions to the microlensing
light curve according to their relative projected proximity to the
lensing mass.  Since this proximity is a function of time, an analysis
of the event light curve allows for the possibility of determining the
surface brightness profile of the source.

The magnification due to a binary lens system depends on additional
parameters relating to the binary system: the mass ratio, the separation
of the lenses, and the angle defining the trajectory of the source relative
to the binary lens orientation.
The magnification due to a binary lens has no simple analytic dependence
on these parameters. However, near the fold caustic, the magnification 
can be approximated by  

\begin{equation}
A_{\rm cau} = A_0 + \frac{b_0}{\sqrt{d}},
	\label{eq:caustic}
\end{equation}

\noindent where $d=d(t)$ is the angular distance (normalized to
$\theta_{\rm E}$) from the caustic to a source element for all such
elements interior to the caustic, $A_0$ is the constant magnification
of the three non-caustic images, and $b_0$ is a scale factor related
to specifics of the lens geometry and the ratio of $\theta_{\rm S}$
to $\theta_{\rm E}$ (e.g., the discussion of Castro \etal\ 2001).
The literature of reports on caustic crossings by stars indicate typical
observed peak magnifications of around 20--30 are common (e.g., Abe \etal\
2003; Afonso \etal\ 2000; Albrow \etal\ 1999; Albrow \etal\ 2000; Albrow
\etal\ 2001a; Albrow \etal\ 2001b).  The sources are generally red giant
stars, with radii of around $R_* \approx 10 R_\odot$.  With $\theta_{\rm
S}/\theta_{\rm E} \sim 10^2$, values of $b_0$ are around 10, and values
of $A_0$ are of order 3--10, as evidenced by the `troughs' between
successive caustic crossings (e.g., Alcock \etal\ 2000).  The particular
form of the approximation in Eq.~\ref{eq:caustic} assumes that the
caustic is a straight line (i.e., the source must be small compared to
the curvature of the caustic) and that the caustic crossing point is
not in the vicinity of a cusp, where the magnification function takes
a different form.  Exterior to the caustic (yet near it so that the
curvature of the caustic can still be ignored), the magnification will
be constant with $A_{\rm cau} = A_0$.  Note that the
time-dependent variation of the polarised flux results entirely from
the second term of Eq.~\ref{eq:caustic} for intrinsically symmetric and
unresolved sources.

\section{Polarization Model}
\label{polmodel}

We review now the intensity profile that will be used with Eq.\
\ref{eq:5} to calculate the microlensing lightcurves.  As stated
above the formalism follows that of Simmons et al.\ (2002).  As
emphasized in that work the model employed is well-suited to evolved
cool stars.  This class of stars exhibits stellar winds that are
significantly stronger than the Sun's, with mass-loss rates ranging
from $10^{-10} M_\odot$
yr$^{-1}$ for typical red giants up to $10^{-5} M_\odot$
yr$^{-1}$ for red supergiant and asymptotic giant branch stars
(e.g., Lamers \& Cassinelli 1999).  The more extreme stellar winds
are clearly dust-driven (e.g., Netzer \& Elitzur 1993; Habing,
Tignon, \& Tielens 1994).  Stars with milder winds are less understood,
possibly driven by a combination of a Parker-style wind augmented
by molecular and dust opacities (Jorgensen \& Johnson 1992).

Assuming the flows are spherically symmetric, the run of the bulk
gas density $\rho$ in the wind with radius $r$ will be given by

\begin{equation}
\rho = \frac{\dot{M}}{4\pi r^2\,v(r)},
	\label{eq:rho}
\end{equation}

\noindent where $\dot{M}$ is the mass-loss rate, and $v(r)$ is the
radial velocity of the flow, beginning subsonically and asymptoting 
to a terminal speed $v_\infty$ at large radius.

The scattering opacity responsible for producing polarisation could
be molecular Rayleigh scattering or dust scattering.  Consequently,
the density profile of the polarigenic species can differ from
Eq.~\ref{eq:rho} by virtue of how molecules or dust are produced
and destroyed as a function of radius.  Particularly in the case
of dust, the location of the condensation radius will be important.
This latter point concerning dust is interesting, since the stars under
consideration have photospheric effective temperatures of 3000--4000~K,
whereas the dust condensation temperature is typically around 1500~K
(Gail \& Sedlmayr 1986).  The implication is that dust-driven winds can
have central cavities (not vacuums, but rather interior zones for which
there is no significant scattering opacity).  Models for dust-driven
winds naturally predict the radial extent of these cavities, yet the
condensation radius is largely unconstrained by observations (Bloemhof \&
Danen 1995).  Microlensing can provide relevant observational constraints
on the location of the condensation radius.  For this pilot study of
the polarimetric signals from caustic crossing events, we choose to
parametrize the scattering number density by a simple power-law with

\begin{equation}
n(r) = n_0 \, \left(\frac{R_*}{r}\right)^\beta,
\end{equation}

\noindent instead of fretting about the details of the wind
acceleration, or how the scattering opacity evolves in the flow.
We additionally assume the envelope is optically thin for this
application, but will explore optical depth effects of the envelope
in a subsequent paper.

We adopt the Stokes vector notation $\mathbf{I} =(I, Q, U, V)^T$
for the intensities, and $\mathbf{F} =(F_{I}, F_{Q}, F_{U}, F_{V})^T$ for the
fluxes.  Following the notation of Simmons \etal\ (2002), the Stokes 
intensities for direct and scattered light will be given by

\begin{eqnarray}
{\bf I}(p,\alpha)&=& {\bf I_{0}}(p)+\frac{3}{16}I_{*} (\beta-1)
\left(\frac{R_{\rm h}}{p}\right)^{\beta-1} \left(\frac{R_{*}}{p}\right)^2\\
\nonumber
& &
\times \tau_{\rm sc}\,g_0(p) \left(
\begin{array}{c} G_{\rm I}\\ G_{\rm P}\cos(2\alpha)\\ -G_{\rm P}\sin(2\alpha) \\ 0   \end{array}
\right)
\label{pm1}
\end{eqnarray}

\noindent where $R_{\rm h}$ is the radius for any `hole' of
scattering opacity, and the total optical depth, $\tau_{\rm sc}$ is
defined as

\begin{equation}
\tau_{\rm sc}=\frac{n_0 \sigma R_{\rm h}}{\beta -1},
\label{eq:tau}
\end{equation}

\noindent with $\sigma$ the scattering cross section, and where the
scatterers are taken to exist only for $r\ge R_{\rm h}$.

For simplicity we shall assume a uniform surface brightness profile
for the star, such that

\begin{equation}
{\bf I_0}(p)= \left\{ \begin{array}{ll}
(I_*,0,0,0) & {\rm for}\;p<R_* \\
0   & {\rm for}\;p \geq R_* \end{array} \right.
\label{pm2a}
\end{equation}

\noindent where $I_*$ is the intensity of the star.  Although we ignore
limb darkening, its main effect would be to enhance the polarisation
from lines-of-sight that are close to the star (bearing in mind that
we are currently ignoring scattering polarisation from the stellar
photosphere, an assumption that we shall justify {\it a posteriori}).
The reason is that at a fixed distance from the star, limb darkening
causes the stellar radiation field impinging upon a given point at this
distance to be more radially directed than is the case for a uniformly
bright star (Cassinelli, Nordsieck, \& Murison 1987).

Also appearing in Eq.~\ref{pm1} is the stellar occultation factor, $g_0$,
that is given by

\begin{equation}
g_0(p)= \left\{ \begin{array}{ll}
1/2 & {\rm for}\;p<R_*, \\
1   & {\rm for}\;p \geq R_*. \end{array} \right.
\label{pm3}
\end{equation}
The occultation factor simply accounts for the fact that radiation
scattered on the far side of the star will not reach the observer.
The integral factors $G_{\rm I}$ and $G_{\rm P}$ are

\begin{eqnarray}
G_{\rm I}&=&\int^{s_{\rm max}}_{0}\sqrt{s^{\beta+1}}
\sqrt{\frac{1-sq}{1-s}}\\
\nonumber
 & &\left\{ \frac{8}{3}\left(\frac{1}{s^2q}\right)
\left[(1-sq)^{-1/2}-(1-\frac{1}{4}sq)\right]-1 
\right\} \mathrm{ds}
\label{pm4}
\end{eqnarray}
and 

\begin{equation}
G_{\rm P}=\int^{s_{\rm max}}_{0} \sqrt {s^{\beta+1}}
\sqrt{\frac{1-sq}{1-s}}ds
\label{pm5}
\end{equation}
with $s=(p/r)^2$, $q=(R_*/p)^2$, $s_{\rm max}=(p/r_{\rm{min}})^2$, for

\begin{equation}
r_{\rm min} = {\rm max}(R_{\rm h},p),
\end{equation}
where $z$ is the line-of-sight coordinate and $r^2=p^2+z^2$.

The fluxes during the microlensing event are then computed from
the integral expressions:

\begin{eqnarray}
F_{\rm I} & = & \int_0^\infty\, \int_0^{2\pi}\,I(p,\alpha)\,
	A_{\rm cau}(d)\,p\,dp\,d\alpha, \\
F_{\rm Q} & = & \int_0^\infty\, \int_0^{2\pi}\,I(p,\alpha)\,\cos 2\alpha\,
	A_{\rm cau}(d)\,p\,dp\,d\alpha, \\
F_{\rm U} & = & \int_0^\infty\, \int_0^{2\pi}\,I(p,\alpha)\,\sin 2\alpha\,
	A_{\rm cau}(d)\,p\,dp\,d\alpha, \\
F_{\rm V} & = & 0.
\end{eqnarray}
	
\noindent As is usual, the observed fractional polarisation can
then be calculated as

\begin{equation}
P=\frac{\sqrt{F_{\rm Q}^2+F_{\rm U}^2+F_V^2}}{F_{\rm I}}
\label{peq}
\end{equation}
where the total flux $F_{\rm I}$ is a sum of the direct stellar flux $F_*$ and the
scattered intensity $F_{I_{\rm sc}}$. The polarisation position angle is
defined as

\begin{equation}
\psi=\frac{1}{2}\tan^{-1}\frac{F_{\rm U}}{F_{\rm Q}}.
\label{pangle}
\end{equation}

\section{Results}

Having described the lensing approximation for a straight fold caustic,
and the underlying source model, we have conducted a parameter study
for microlensing light curves associated with caustic crossing events as
various source and lens properties are varied.  In displaying results,
our goal is to highlight those features that pertain to elucidating
the properties of the source; therefore, instead of light curves as
a function of time, we choose to plot flux observables as a function
of caustic position relative to the star.  To do so, we define the
coordinate $x_{\rm lens}$ as the normal projected distance between the
straight line caustic and star centre in the source plane.  Then $y_{\rm
lens}$ is the coordinate along the caustic direction.  Both coordinates
are normalized to the stellar radius $R_*$.  The case $x_{\rm lens} < 0$
is when the star centre lies inside the caustic; the case $x_{\rm lens}
> 0$ is when the star centre lies outside the caustic; and $x_{\rm lens}
= 0$ is the moment of transit for the star centre.

Conversion to a time coordinate is achieved with $t = (x_{\rm lens}
\,R_*)/(D_{\rm S} \,\mu_{\rm rel,x})$, where $\mu_{\rm rel,x} = \mu_{\rm rel}
\cos \gamma$, with $\mu_{\rm rel}$ the magnitude of the relative proper
motion between the source and caustic, and $\gamma$ is the trajectory
orientation of the source relative to the caustic.  So $\gamma=0^\circ$
means the source is traveling in the $+x$ direction in the frame of the
caustic, whereas $\gamma= \pm 90^\circ$ means the source is moving in the
$\pm y$ direction, respectively.  The orientation of the trajectory has
no bearing on the shape of the lensing light curves in time, except as a
`stretch' factor for scaling purposes; for example, $\gamma$ does not
affect the value of the peak polarisation achieved during the event.

\begin{figure*}
\centerline{\epsfig{file=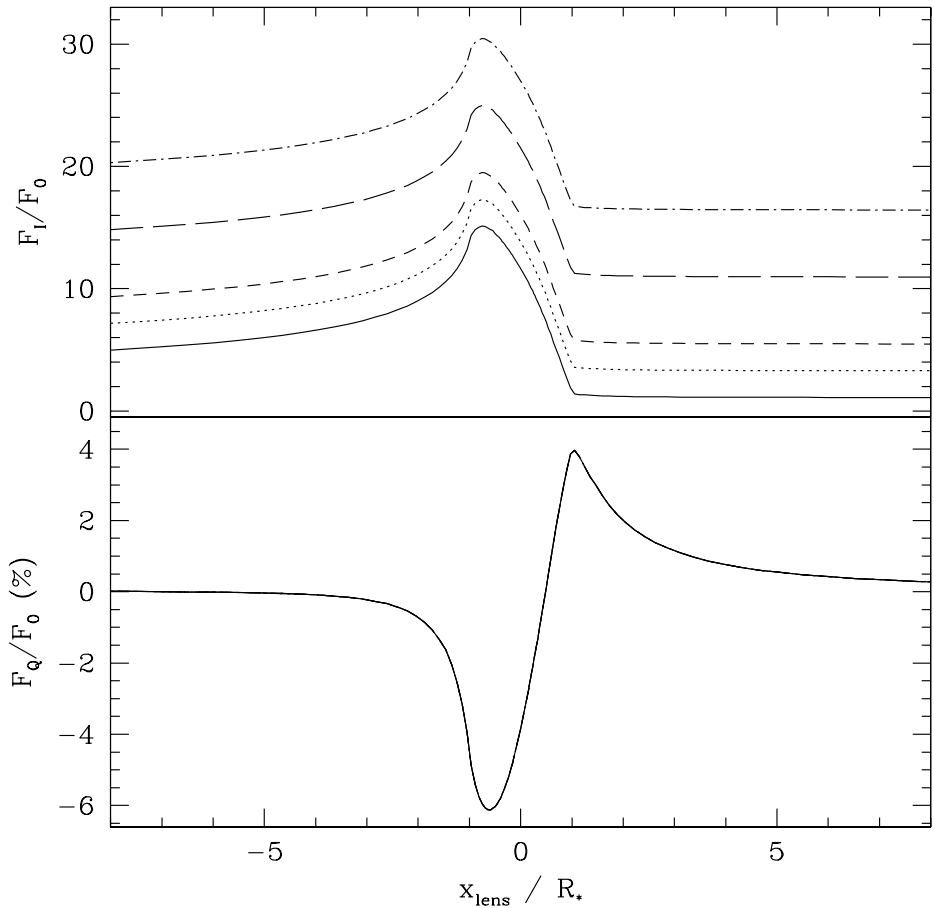,angle=0,width=13cm}}
\caption{\small{Plots of the variation in total flux (upper panel) and
the normalized polarised flux (lower panel) for a scattering envelope during
a fold caustic crossing event.  The lower axis displays the location
of the caustic in the source plane (normalized to $R_*$) with respect
to the centre of the star, with negative values for when the star lies
{\em interior} to the caustic, and positive values for when it lies
{\em exterior}.  Values of $\beta=2$, $R_{\rm h}=R_*$
(no cavity), and an envelope optical depth of $\tau_{\rm sc} = 0.1$
are adopted.  A lens parameter of $b_0=10$ is fixed, but $A_0$
takes values of $1, 3, 5, 10,$ and 15
(in order of increasingly stronger peak fluxes).  As described in the
text, variations in $A_0$ change the total flux light curves, but not
the polarisation, because $A_0$ is a constant of the integration
on either side of the fold caustic.  It is $b_0$ and the properties of
the envelope that set the shape and amplitude of the polarimetric variations.
\label{plot0} }}
\end{figure*}

\subsection{Test Case}

Figure~\ref{plot0} shows model results for a test scenario.  The
upper panel shows the flux of the source, relative to the case of
being unlensed, as a function of the caustic location.  The lower
panel is for the polarised flux, also normalized to the unlensed
flux of the source $F_0$, and multiplied by 100 to simulate a kind
of percent polarised flux (the reason for this rather odd choice
will become apparent in a moment).  The envelope parameters are
fixed with an envelope optical depth of $\tau_{\rm sc}=0.1$,
density parameter $\beta =2$, and no cavity (i.e., $R_{\rm h} =
R_*$).  The lens parameter $b_0$ is fixed at a value of 10, but
$A_0$ is allowed to vary from 1 to 15, with larger values giving
stronger peak flux magnifications.

We have argued that $A_0$ should have no bearing on the polarised
emission, and indeed that is seen to be the case in Figure~\ref{plot0}.
All of the variation of the polarized flux comes from the second term in
Eq.~\ref{eq:caustic} with scaling $b_0$.  In the standard representation
of polarisation with $p = F_{\rm Q}/F_{\rm I}$.  The fractional or percent
polarisation can be affected by the value $A_0$, because the total flux
variation as the microlensing event evolves does depend on $A_0$.  A value
of $A_0=1$ has been adopted for the rest of the model calculations.

\begin{figure*}
\centerline{\epsfig{file=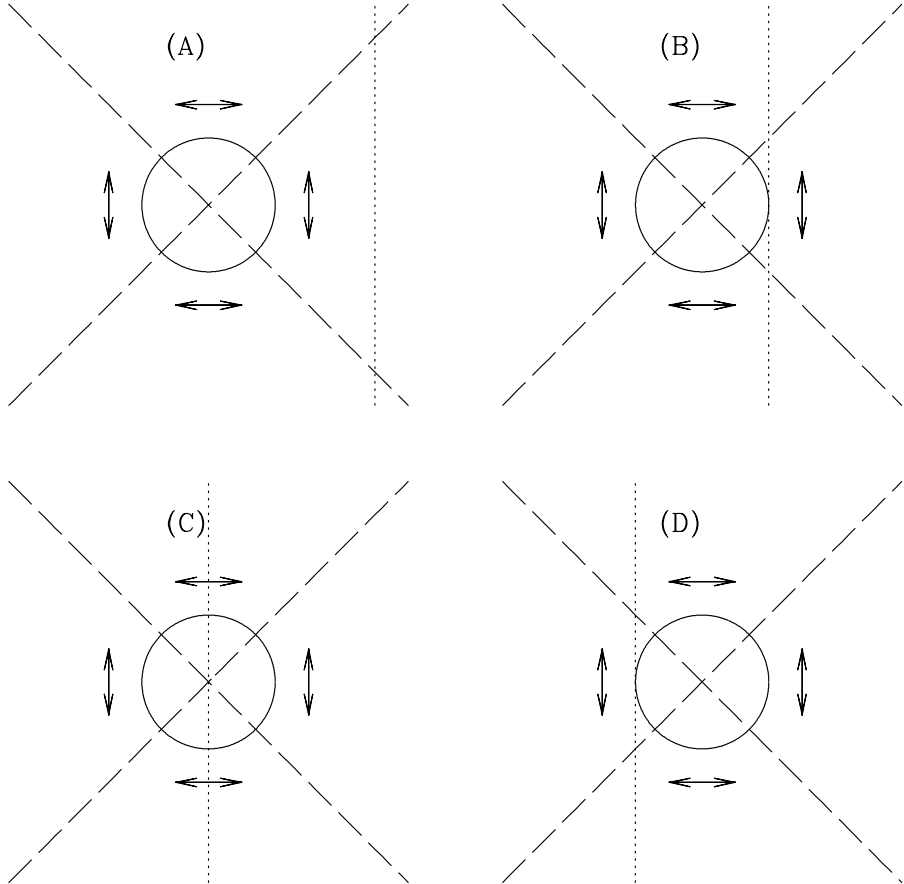,angle=0,width=13cm}}
\caption{\small{
Schematic of the geometry for the transit of the star and circumstellar
envelope across a fold caustic.  Shown are four times (A) -- (D).
The star is cross-hatched.  The vertical dotted line is the caustic,
with left being interior and right being exterior.  The double-headed
arrows are representative polarisation vectors at those points
(vertical being $Q>0$ and horizontal being $Q<0$).  Finally, the
diagonal dashed lines are the locus of points where the emergent
$Q$-intensity would be zero; these then are `null' lines where
$Q$ switches sign, and the polarisation position angle rotates
$90^\circ$ between adjacent zones.
\label{plot0b} }}
\end{figure*}

It is useful at this point to introduce a schematic figure that
demarcates regions contributing to the polarised light as the caustic
crossing evolves.  Figure~\ref{plot0b} shows four panels, with the
source moving from inside the caustic to outside in the sequence (A) to
(D).  The vertical dotted line is the caustic, hence the left region is
interior to the caustic and the right region is outside it.  The star
is the cross-hatched region.  The double-headed arrows show the sense of
orientation that would result for the emergent polarised flux from right
angle scattering in the plane of the sky if the source were resolved.
The dashed diagonal lines are where $Q=0$.  Thus the figure is useful
in mapping how the different zones of polarised flux will contribute
to the total polarised emission from the unresolved lensed source as
a function of its location relative to the caustic line.

The variation of the model polarised flux shows interesting sign changes.
Here we are assuming an observational scenario in which a second
exiting caustic crossing has been predicted from an earlier interior
crossing event.  Thus in Figure~\ref{plot0b}
at time (A), the source is approaching the caustic to exit.
The quadrant closest to the fold is dominated by scattered light with
$Q>0$, and so initially the polarisation is oriented parallel to the
fold caustic in the sky.  As the event progresses, the situation of case
(B) is reached.  Only scattered light leftward of the caustic will be
subject to a magnification {\em gradient}, thus breaking the symmetry
and leading to a net observed polarisation.  By time (B), the
light being most strongly magnified is for $Q<0$, and so the polarisation
position angle has rotated $90^\circ$ by the time the caustic is first
tangent to the photosphere.  By time (C), a minimum in the polarised flux
has passed.  Now the polarisation changes rapidly, so that by the time
the caustic is tangent to the far side of the photosphere, the polarised
flux has changed sign again, becoming positive.  Note in terms of the
polarised light $F_{\rm Q}/F_{\rm I}$, the percent polarisation will rise
significantly, because the magnification of the stellar photosphere is
minimized once the photosphere has completely exited the caustic.

\subsection{Variable Envelope Optical Depth}

\begin{figure*}
\centerline{\epsfig{file=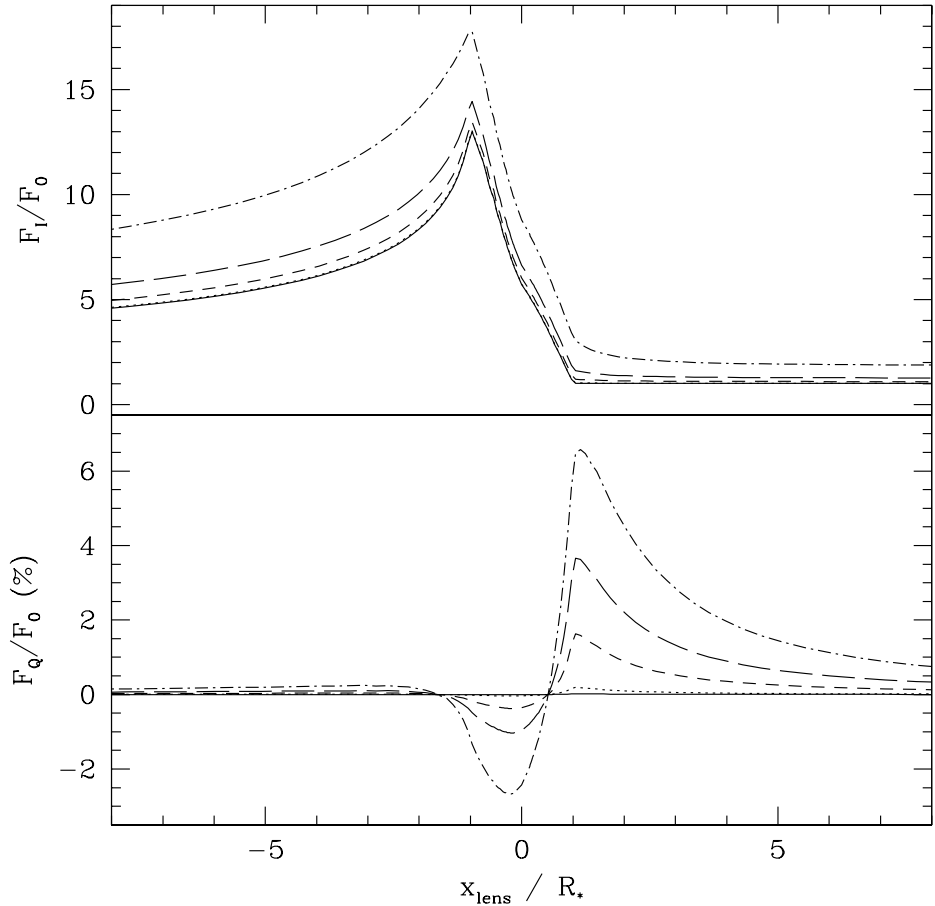,angle=0,width=13cm}}
\caption{\small{Plots like Fig.~\ref{plot0}, but now with fixed
lens parameters and variable envelope optical depths, and with the
lower panel as percent polarisation (normalized by the total intensity
flux $F_{\rm I}$).  In this case values of $\beta=2$, $R_{\rm h}=R_*$
(no cavity), $A_0=1$, $b_0=10$ are assumed.  The different curves
are for different envelope scattering optical depths, with $\tau_{\rm
sc} = 0.001, 0.01, 0.1, 0.3,$ and 1.0 (in order of increasingly
stronger peak polarisations).  Note the change in sign of the
polarisation as the stellar photosphere transits the caustic, and
the strong peak value that results immediately after the photosphere
has completely exited the caustic.  \label{plot1} }}

\end{figure*}

Figure~\ref{plot1} shows the response of the flux and polarisation light
curves to different values of envelope optical depth.  The lower panel
now plots the standard form of polarisation, $p=F_{\rm Q} /
F_{\rm I}$, and hereafter.  Values of $\tau_{\rm sc} = 0.001, 0.01, 0.1,
0.3,$ and 1.0 are used.  A value of $A_0=1$ and $b_0=10$ are used with
a density distribution described by $\beta=2$.  These values for $A_0$,
$b_0$, and $\beta$ will be standard in our calculations unless noted
otherwise.  The envelope has little effect on the flux light curve except
when $\tau_{\rm sc} \gtrsim 0.1$.  On the other hand, the amplitude of
the polarisation light curve is approximately linear in $\tau_{\rm sc}$.

The envelope is reasonably thin, and so the flux magnification is
dominated by the photospheric emission.  Agol (1996) has investigated
the polarimetric variations from scattering polarisation in stellar
atmospheres.  We have purposely ignored photospheric contributions
to the polarised emission, because circumstellar envelopes are more
efficient at producing polarised emission (albeit, this is a function
of optical depth), and because we wish to investigate the effects of a
circumstellar envelope for the light curves.  Although $A_0$ and $b_0$
are not exactly the same, the case of the solid line in the upper panel
of Figure~\ref{plot1} is roughly comparable to the $r=0.01$ case shown in
the upper panel of Agol's Figure~1.  

Overall, the polarisation curves are generally similar to those of
Agol (1996); however, there are notable quantitative and qualitative
differences.  First quantitatively, the peak polarisation achieved by
photosphere crossings were rarely in excess of 1\%, and in some cases
only a few tenths of a percent, whereas thin scattering envelopes can
achieve values in excess of 5\% when $\tau_{\rm sc}$ is large enough
($\tau_{\rm sc}\gtrsim 0.5$).

There are qualitative differences as well.  The underlying source
models are drastically different.  For example, a stellar photosphere
has an intrinsic polarisation profile (as a function of $p$) that
is maximum at the stellar limb and decreases to zero at the centre
of the star.  In our case the photospheric polarisation is ignored
because it is small compared to the envelope polarisation.  The
extended envelope has a polarisation profile that is zero at large
distance from the star, and increases toward the stellar limb.
The polarisation peaks outside the stellar limb, and decreases to zero
again at the stellar centre.  So the source models are quite different,
but at the same time the variable polarisation from microlensing for the
two cases are somewhat similar.  Basically, the photospheric polarisation
has a discontinuous jump in moving from off the star across the stellar
limb, whereas the circumstellar envelope has a more gradual peak off the
stellar limb.  Microlensing involves a weighted surface integral, thus
`smoothing' over these detailed differences, leading to somewhat similar
light curves (Dominik 2004a).  Still, the fact that the peak polarisation
appears at the stellar limb for a photosphere in a discontinuous way
explains why there is a peak (sometimes cuspy) at $x_{\rm lens} = -
R_*$ in Agol's models and not in ours.  (This is $y_* = - r$ in Agol's
notation; see his Fig.~1.)

\subsection{The Influence of a Cavity}

\begin{figure*}
\centerline{\epsfig{file=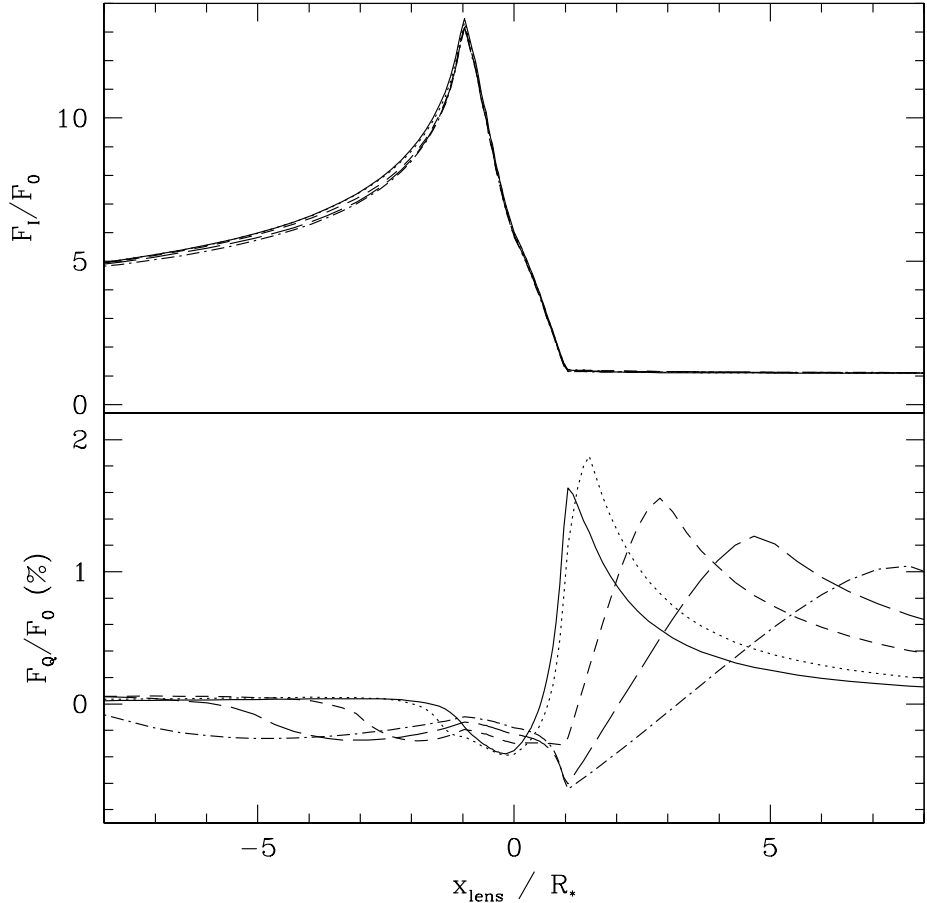,angle=0,width=13cm}}
\caption{\small{Illustration of how a cavity of scattering opacity impacts
the polarisation variation.  All of the curves are for $\tau_{\rm sc}
= 0.1$.  Each is distinguished by the extent of the cavity, with $R_{\rm
h} = 1.0, 1.5, 3.0, 5.0,$ and $8.0 R_*$, with larger cavities yielding
peak polarisations at larger values of $x_{\rm lens}$.  Note that the
total flux curve (upper panel) is little influenced by the cavity extent
because the scattering envelope is optically thin.  Although basically
similar, there are some notable qualitative differences in the variation
of polarisation when a cavity is present (see Fig.~\ref{plot3}).
\label{plot2}
}}
\end{figure*}

Figure~\ref{plot2} shows how a cavity of scatterers at the inner
envelope impacts the polarisation light curves.  The different
curves are for different `hole' radii of $R_{\rm h} = 1.0, 1.5,
3.0, 5.0$ and $8.0 R_*$.  Clearly, as the extent of the cavity
increases, the polarimetric variation occurs over a longer time-scale,
with the peak polarisation shifting toward larger values of positive
$x_{\rm lens}$, and the negative `trough' growing in extent toward
negative $x_{\rm lens}$ as a precursor to the transit of the star.

In each case the envelope optical depth is maintained at $\tau_{\rm
sc} =0.1$.  First, this is rather optically thin, as evidenced by the
total flux light curve (upper panel) which does not vary much between the
different cases, and is dominated largely by the photosphere.  Second,
the scale of the polarisation is to zeroth order determined by the value
of $\tau_{\rm sc}$, which explains why all of the curves have similar
peak polarisation values.  Third, holding $\tau_{\rm sc}$ constant
implies conserving the total number of scatterers; hence although the
different cases shown have the same density distribution at $\beta=2$,
these require different density scales $n_0$, so as to maintain fixed
$\tau_{\rm sc}$.  Using Eq.~(\ref{eq:tau}), the density scale for
fixed envelope optical depth is given by

\begin{equation}
n_0 = (\beta-1)\,\frac{\tau_{\rm sc}}{\sigma\,R_{\rm h}}
\end{equation}

\noindent So our approach for inserting a cavity is not to delete
scatterers, but to redistribute them outward.
The goal of considering cavities is to illustrate how microlensing
can neatly trace the cavity extent through the polarisation light
curve, which is relevant for the case of red giants that can form dust
in their winds at a condensation radius that is offset from the
stellar photosphere.  

\begin{figure*}
\centerline{\epsfig{file=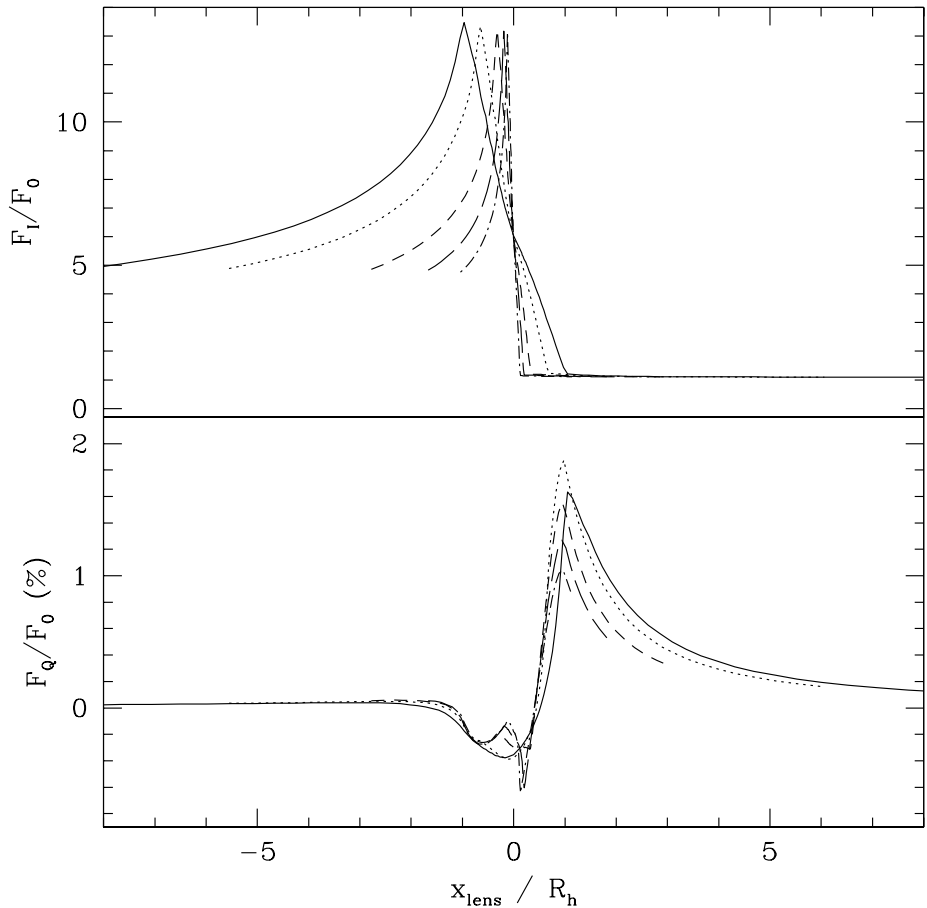,angle=0,width=13cm}}
\caption{\small{The results shown in Fig.~\ref{plot2}, except now the
lower axis for the position of the fold caustic is normalized to the
size of the cavity $R_{\rm h}$.  From Fig.~\ref{plot2}, the total flux
variation was set by the passage of the photosphere across the caustic.
The polarimetric variations on the other hand are determined the size of
the cavity.  Although the positive peak polarisation has variable width,
the negative `trough' has approximately constant width.  The solid line
is for no cavity (i.e., $R_{\rm h}=R_*$); notably the presence of a cavity
changes the qualitative shape of the trough, with some recovery toward net
zero polarisation followed by a sharp drop toward more negative values.
\label{plot3} }} 
\end{figure*}

Figure~\ref{plot3} shows the same curves of Figure~\ref{plot2}, but with 
$x_{\rm lens}$ normalized to $R_{\rm h}$ instead of $R_*$, which nicely
shows how the polarimetric variations are set by the crossing of the cavity.
Clearly, the peak polarisation is affected by the cavity (generally smaller),
and substructure is seen around the passage of the photosphere from inside
the caustic to outside.  The peak polarisation consistently occurs just before
the cavity transits entirely out of the caustic, and so becomes an excellent
tracer of the cavity extent relative to the stellar radius, which can be
determined from the total flux variations.  

In other words the total flux variations show a peak at a time when
the stellar limb just begins to transit the caustic.  As the event
proceeds, the total flux shows a precipitous drop, and then goes
flat.  The extended envelope can produce a tail of enhanced brightness
after the star has transited out of the caustic, but the drop is
dominated by the star.  So that time-scale is $2t_* = 2\theta_*/\mu_{\rm
rel}$, where $\theta_*$ is the angular radius of the star.  On the
other hand, from this time until the peak polarisation is achieved
will require a time $t_{\rm h} = \theta_{\rm h}/\mu_{\rm rel}$.
Consequently $R_{\rm h} / R_*$ will be given by $t_{\rm h}/t_*$.

\subsection{The Impact of the Density Distribution}

\begin{figure*}
\centerline{\epsfig{file=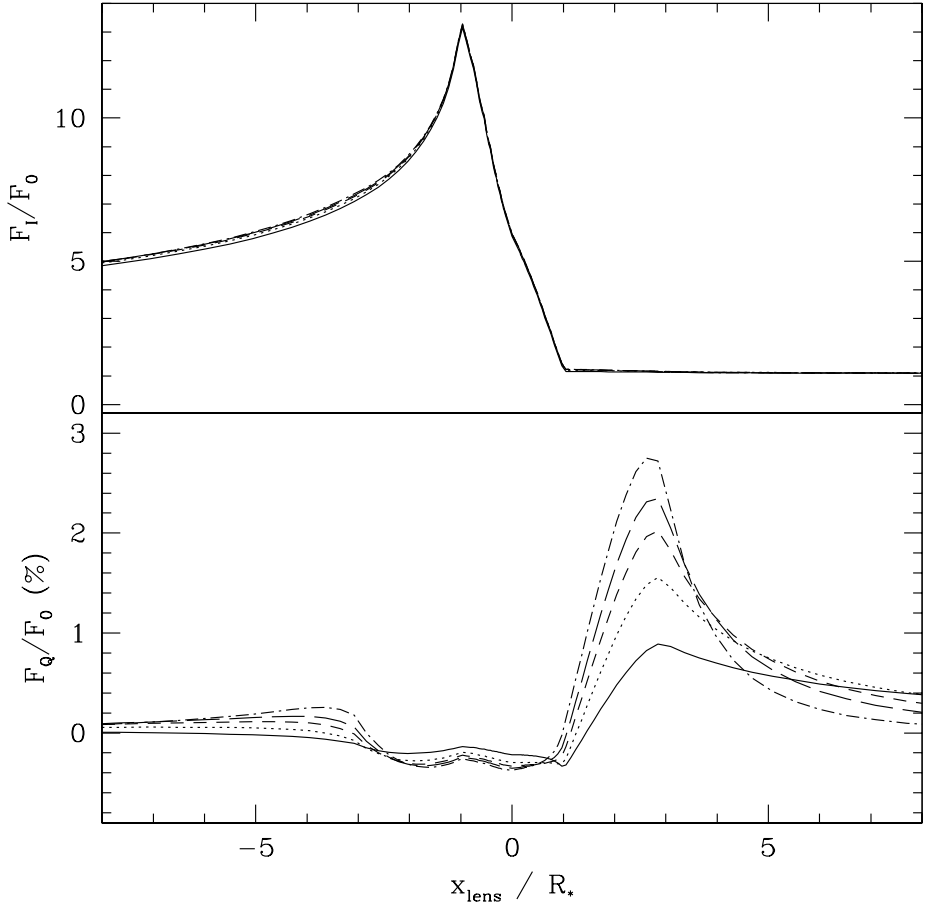,angle=0,width=13cm}}
\caption{\small{Here only the value of $\beta$ is allowed to vary.
The flux light curve hardly changes, whereas the polarimetric profiles,
although similar qualitatively, are seen to vary significantly in
amplitude.  The values of $\beta$ are 1.5, 2.0, 2.5, 3.0, and 4.0,
in order of stronger positive peak polarisations, and the envelope
sports a cavity with $R_{\rm h}=3.0R_*$.  \label{plot4}}}

\end{figure*}

Models for the flux and polarisation variations have been generated
for different envelope density distributions, and the results are
displayed in Figure~\ref{plot4}.  The curves are for $\beta=1.5,
2.0, 2.5, 3.0$ and 4.0.  A cavity with $R_{\rm h}=3.0R_*$ has been
adopted.

The different cases are all for a fixed value of $\tau_{\rm sc}=0.1$.
As noted before, this is in essence achieved via redistribution of
scatterers.  In this case there is a fixed hole.  Changing $\beta$
makes the density distribution more or less steep.  As $\beta$ is
made to increase, keeping $\tau_{\rm sc}$ fixed results in increased
values of $n_0$, and so the peak polarisation that is dominated by the
number density of scatterers at the limb of the cavity increases as well.

So $\beta$ does not necessarily lead to larger polarisations; here
it is an artefact of maintaining a constant value of $\tau_{\rm sc}$.
Fundamentally what $\beta$ does is to alter the slope of the polarisation
curves after the cavity has completely passed out of the caustic,
making the slopes steeper with increasing $\beta$.  In fact, in the
limit that the star can be treated as a point source, the asymptotic
slope of the polarisation for relatively large values of $x_{\rm lens}$
can be derived analytically.  The derivation is found in the Appendix;
here just the result is quoted.  Asymptotically, the polarised flux
({\it not} the percent polarisation) will be given by

\begin{equation}
F_{\rm Q} \propto \left(\frac{R_{\rm h}}{x_{\rm lens}}\right)^{(2\beta-1)/2}
	\propto t^{-(2\beta-1)/2}.
	\label{eq:latetime}
\end{equation}

\noindent So larger values of $\beta$ lead to steeper declines in the
polarised flux as the microlensing event progresses.  Two points should
be mentioned.  First, although real sources may not follow a power-law
distribution for the density of scatterers in portions of their extended
envelopes, a value of $\beta=2$ is reasonable to expect at large scale
for a spherical wind flow, for which case $F_{\rm Q} \propto t^{-3/2}$.
Second, the preceding equation is only valid both for when the star
can be treated as a nearly point source of illumination with respect to
the scattering envelope and when the caustic can be approximated as a
straight line.  Even if the asymptotic trend of Eq.~(\ref{eq:latetime})
is not achieved in a real event, it does provide useful insight and
limiting behavior for the modeling effort.

\subsection{Variation of the Magnification Gradient}

\begin{figure*}
\centerline{\epsfig{file=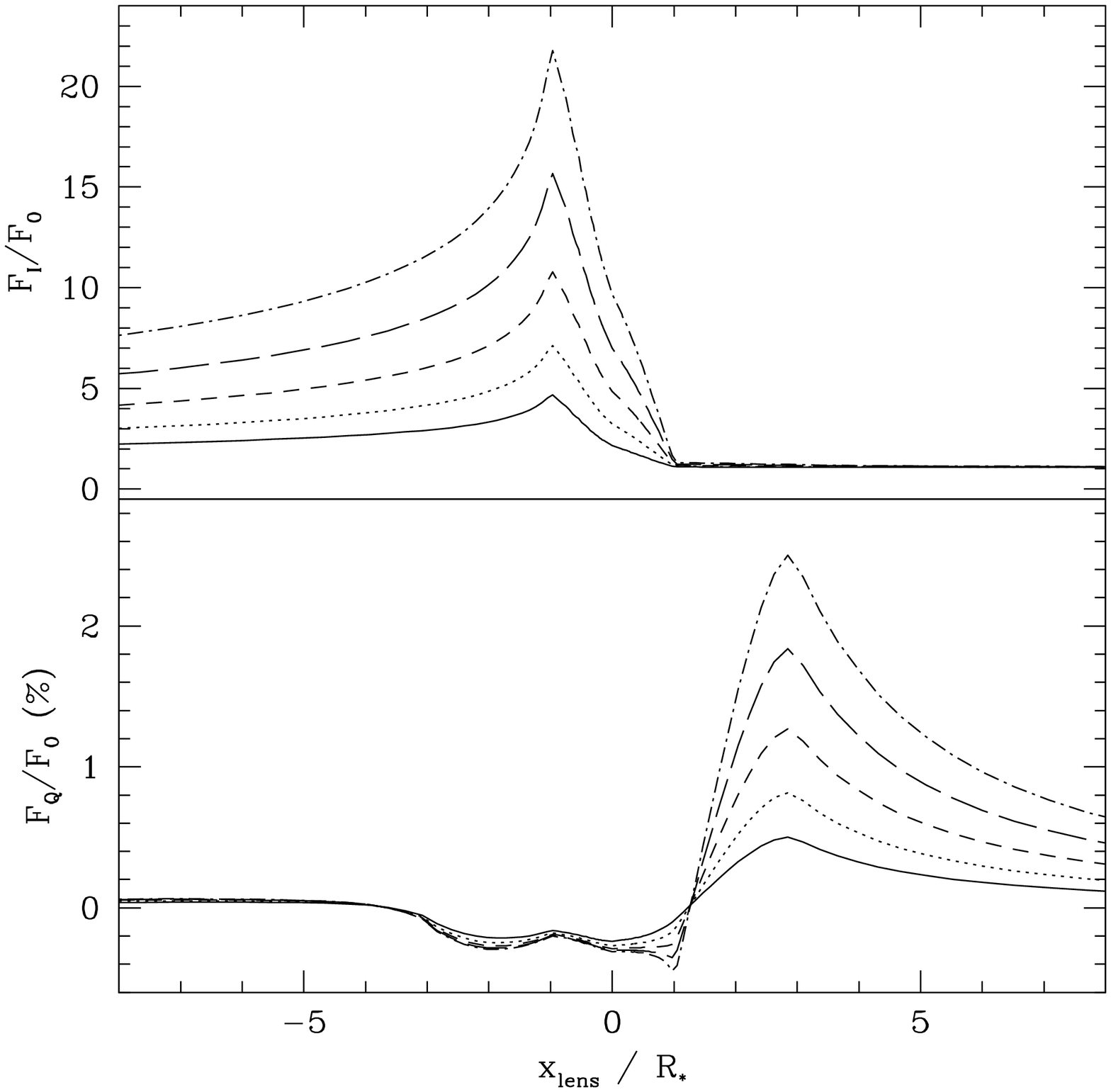,angle=0,width=13cm}}
\caption{\small{These figures are for fixed envelope properties, but for
different values of the lens parameter $b_0 = 3, 5, 8, 12,$ and 17 and $A_0$
fixed at unity.  Both the strength of the peak flux and peak polarisation
vary almost linearly with the value of $b_0$.  Variations in $A_0$ would have
no effect on the polarisation, but would influence the total flux curves.
\label{plot5}}}
\end{figure*}

Figure~\ref{plot5} shows model results as the value of $b_0$ is
varied, with values of $3, 5, 8, 12,$ and 17 (with $A_0=1$ fixed).
With $b_0$ relatively large compared with $A_0$, the polarisation
varies little for $x_{\rm lens} < 1$, prior to when the star has
passed out of caustic.  The reason is that the polarisation
is a ratio of the polarised light to the total flux, and both
scale as $b_0$.  Clearly, the total flux shown in the upper panel
is strongly dependent on $b_0$, increasing essentially linearly
with $b_0$.  Similarly, the polarisation {\it after} the star
has passed out of the caustic is affected by $b_0$, because
now $F_{\rm I}$ is a constant that depends primarily on
the photospheric flux multiplied by $A_0$, whereas $F_{\rm Q}$
still depends on the value of $b_0$.

\section{Discussion}

This study has demonstrated that polarimetric observations of caustic
crossing events could be used to probe circumstellar envelopes.
Such data could constrain the number density of the scatterers
within the envelope, detect the presence and trace the extent of a
central cavity around the source photosphere, and provide information
about the density distribution of scatterers.

To summarize, a comparison of the time-scale for the total flux
variations against that of the polarimetric variations yields the
extent of the cavity relative to the stellar radius.  In the case
of dust producing cool star winds, that information is sufficient
to test models that predict the location of the dust condensation
radius.  The late time evolution of the polarised flux is set by
the density distribution (our $\beta$-value for this work).
Lens parameters $A_0$ and $b_0$ can be derived from fitting the
variation of the photospheric flux during the caustic crossing,
in which case model fits to the polarisation level will give the
value of $\tau_{\rm sc}$ for the circumstellar envelope.

To connect our model results with applications to observed events, we must
relate the envelope properties for the model to those of real sources.
Some help toward this end is provided by Netzer \& Elitzur (1993), who
describe model results for dust-driven winds.  In their Eq.~(10),
they provide an expression relating stellar parameters $\dot{M}$,
$v_\infty$, and $L_*$ to the flux mean optical depth of the envelope
$\tau_{\rm F}$:

\begin{equation}
\frac{\dot{M}}{2\times 10^{-5} \,M_\odot\,\;{\rm yr}^{-1}} = \tau_{\rm F}\,
	\frac{L_*/10^{4}L_\odot}{v_\infty/10\,{\rm km\;s}^{-1}},
\end{equation}

\noindent here shown in slightly modified form from their paper.
The value of $\tau_{\rm F}$ will not equal the value of $\tau_{\rm sc}$
that we use to characterize our models; however, the flux mean opacity
does give an overall scale related to the optical depth of the envelope.
Although relating $\tau_{\rm sc}$ to $\tau_{\rm F}$ will depend on the
particular opacities involved, one might generally expect that the two
will scale together.  The models of Netzer \& Elitzur show that the
minimum mass-loss for dust driving to be dominant is around $10^{-7}
\,M_\odot\,\;{\rm yr}^{-1}$.  At this value for a star with $L_*=10^4
L_\odot$ and $v_\infty = 10$ km s$^{-1}$, the flux mean optical depth
will be 0.02 (of course, the optical depth at a wavelength of interest
can be higher or smaller). 

One of the challenges in detecting polarized signals in real events,
such as this hypothetical Bulge star, is that the crossing of the
caustic by the photosphere will typically take only a few hours (e.g.,
Alcock \etal\ 2000).  We can estimate the detectability of polarizations
predicted by our models.  Using a Kurucz model for a cool subgiant of
$g=3.5$ and $T=4500$~K (parameters similar to OGLE-1999-BUL-23 from
Albrow \etal\ 2001), the I-band flux at a distance of 8 kpc is estimated
to be $3\times 10^{-13}$ erg s$^{-1}$ cm$^{-2}$, with a corresponding
magnitude of about $m_I\approx 18$.  Of course, during the lensing event,
the source brightens, and magnifications by an order of magnitude are
typically achieved, at which point $m_I \approx 15.5$.  The time $t$
required to achieve a given signal-to-noise ratio $S/N$ for a telescope
of diameter $D$ in centimeters, allowing for Poisson noise only, will be

\begin{equation}
t \approx \,9\frac{(S/N)^2}{D^2}.
\end{equation}

\noindent Our models indicate that peak polarizations of about 1\%
will be achievable during caustic crossings.  A $5\sigma$ detection at
this polarization level requires $S/N = 500$.  Additionally, exposures
are needed at 8 position angles in order to construct the Stokes $I$,
$Q$, and $U$ fluxes (eight to eliminate systematics).  Consequently, the
required exposure time in total for this detection level, not counting
overhead, will be a little over 30 minutes using a 1-meter telescope.
Although this exposure estimate is a lower limit (owing to neglect of
background, inefficiencies, and telescope overheads), the required
exposure is about 10\% of the duration of the photosphere crossing,
even smaller for the bulk of the circumstellar scattering envelope,
and can be reasonably obtained with modest facilities equipped with
polarimetric instrumentation. \\

Although the original goal of microlensing surveys was to deduce the
properties of dark matter in the Milky Way, it is clear that a vast
range important of by-products have resulted from the survey effort, from
catalogs of variable stars to observations of finite source effects (as
described in the Introduction).  Our contribution to the topic of finite
source effects has been to point out how novel and valuable information
about circumstellar envelopes might be obtained through polarimetric
monitoring of events involving binary lenses and sources that may have
substantial winds.

Certainly, our models include some simplifying assumptions, such as
ignoring polarisation from the photosphere and the effect of limb
darkening.  Neither of these are severe; indeed, both will tend to
{\it increase} the peak polarisations above those predicted by our
models.  Photospheric polarisation is expected to be smaller than the
circumstellar contribution for stars with significant winds, but its
contribution should add to the $Q$ and $U$ fluxes constructively, and not
destructively.  Limb darkening will tend to mollify the effects of the
finite depolarisation factor, thereby increasing the peak polarisation
from the inner wind where the density of scatterers is larger (although
limb darkening will have little or no impact in the case of significant
central cavities).  We have also not allowed for optical depth effects,
by way of multiple scattering effects and extinction of the photospheric
emission.  We intend to consider these effects in a separate paper.
However, substantial scattering optical depths will be more important for
quite dense winds, like those of red supergiants and asymptotic giant
branch stars.  For the more common red giant stars, the optically thin
assumption will be a good assumption. \\

\section*{Acknowledgements}

The authors are grateful to an anonymous referee who made critical comments
that led to needed clarifications of the text.  This research was supported
by a grant from the National Science Foundation (AST-0354262).


\appendix

\section{Obtaining $\beta$ from the Asymptotic Decline of the Stokes $Q$-Flux}

Consider a source that lies inside the caustic and is passing out
of it.  (The arguments that follow also hold true when the source
first approaches the caustic.) After the photosphere of the star
has completely passed out of the caustic, the polarisation achieves
a strong positive (by our convention) peak value.  This occurs
because only the scattering envelope lies interior to the caustic
and so is subject to the selective amplification, whereas the
photosphere is amplified by an approximately constant 
value.  As the event evolves, more of the envelope transits the
fold caustic, and the polarised signal at a given moment is given
primarily by the highest value of the polarised flux along the caustic
itself.  The scale of this polarisation is set by the scattering
optical depth of the envelope, but the slope of the polarised flux
light curve is determined by the density distribution of scatterers.
Here, we derive this relation between the lens position and the 
dependence of the polarisation on $\beta$.

As the star moves farther along from the caustic, the scattered light is
accurately described by a point source.  Equation~(\ref{pm1})
indicates that far from the star, $Q(p,\alpha) \approx Q_0 \, (R_{\rm
h}/p)^{\beta+1} \cos 2\alpha$.  The polarised flux will be given by

\begin{eqnarray}
F_{\rm Q} & = & \frac{R_*^2}{D^2}\,\int_0^\infty \, dx\, A(x)\,\int_{-\infty}^\infty\,dy \,Q(p,\alpha)
	 \nonumber \\
 & = & Q_0\,\frac{R_*^2}{D^2}\,\int_0^\infty \, dx\,\frac{b}{\sqrt{x}}\, 
	\int_{-\infty}^\infty\,dy\,
	\frac{(x+x_l)^2-y^2}{(x+x_l)^2+y^2}
	\nonumber \\
 & & \times \, \left[\frac{R_{\rm h}^2}{(x+x_l)^2+y^2}
	\right]^{(\beta+1)/2}.
\end{eqnarray}

\noindent Factoring out $(x+x_l)$ and making a suitable change of
variable, the integration in $y$ can be evaluated numerically for any
particular value of $\beta$.  For $\beta$ an integer, analytic integration
formulae will apply.  Because the result is not critical for our concerns,
we simply denote the result of the $y$ integral as $Y(\beta)$, giving

\begin{equation}
F_{\rm Q} = Q_0\,b\,R_{\rm h}^{\beta+1}\,\frac{R_*^2}{D^2}\,Y(\beta)\,
	\int_0^\infty\,\frac{dx/\sqrt{x}}{(x+x_l)^\beta}.
\end{equation}

\noindent Using a substitution with $z = \sqrt{x/x_l}$, the integral will
have a similar form as the one for $y$, but with a different dependence
on $\beta$.  We define the result of this integration to be $X(\beta)$,
leading to

\begin{equation}
F_{\rm Q} = Q_0\,b_0\,\left(\frac{R_{\rm h}}{x_l}\right)^{(2\beta-1)/2}\,R_{\rm h}^{3/2}\,
	\frac{R_*^2}{D^2}\,X(\beta)\,Y(\beta).
\end{equation}

Observationally, the lens position is linear with time $t$, and so
$\log F_{\rm Q} = -\frac{2\beta-1}{2}\log t + \log t_0$, where the other
factors have been collected into the variable $t_0$.  Of course this is a
power-law, and its slope is directly related to the value of $\beta$ for
the density distribution of scatterers in the envelope.  For example, many
of our models employ $\beta=2$, in which case the asymptotic polarised
flux varies as $F_{\rm Q} \propto t^{3/2}$.  It should be pointed out
that this limiting behavior will only be achieved in the tail of the
polarised light curve, somewhat following the polarimetric peak.

\end{document}